\documentclass[%
 aip,
 amsmath,amssymb,
preprint,
]{revtex4-1}

\usepackage{graphicx}
\usepackage{dcolumn}
\usepackage{bm}
\usepackage[utf8]{inputenc}
\usepackage[T1]{fontenc}
\usepackage{mathptmx}

\usepackage{amsmath}
\usepackage{hyperref}
\usepackage{tikz}

\usepackage{tikz}
\usepackage{pgfplots}
\usetikzlibrary{patterns}
\usetikzlibrary{plotmarks}
\usetikzlibrary{positioning}
\usetikzlibrary{external}
\usetikzlibrary{patterns}
\tikzset{external/system call={pdflatex \tikzexternalcheckshellescape -halt-on-error 
-interaction=batchmode -jobname "\image" "\texsource" && 
pdftops -eps "\image".pdf}}
\tikzsetexternalprefix{external_figs/}

\begin{document}


\title[Charge carrier density, mobility and Seebeck coefficient of melt-grown bulk \hbox{ZnGa$_2$O$_4$} single crystals]{Charge carrier density, mobility and Seebeck coefficient of melt-grown bulk \hbox{ZnGa$_2$O$_4$} single crystals}

\author{Johannes Boy}
\email[]{boy@physik.hu-berlin.de}
\affiliation{Novel Materials Group, Humboldt-Universität zu Berlin, Newtonstraße 15, 12489 Berlin, Germany}
\author{Martin Handwerg}
\affiliation{Novel Materials Group, Humboldt-Universität zu Berlin, Newtonstraße 15, 12489 Berlin, Germany}
\author{R\"udiger Mitdank}
\affiliation{Novel Materials Group, Humboldt-Universität zu Berlin, Newtonstraße 15, 12489 Berlin, Germany}
\author{Zbigniew Galazka}
\affiliation{Leibniz-Institut f\"ur Kristallz\"uchtung, Max-Born-Strasse 2, 12489 Berlin, Germany}
\author{Saskia F. Fischer}
\email[]{sfischer@physik.hu-berlin.de}
\affiliation{Novel Materials Group, Humboldt-Universität zu Berlin, Newtonstraße 15, 12489 Berlin, Germany}

\date{\today}

\begin{abstract}
The temperature dependence of the charge carrier density, mobility and Seebeck coefficient of melt-grown, bulk \hbox{ZnGa$_2$O$_4$} single crystals was measured between 10 K and 310 K. The electrical conductivity at room temperature is about $\sigma=286$ S/cm due to a high electron concentration of $n=3.26\cdot10^{19}$ cm$^{-3}$, caused by unintenional doping. The mobility at room temperature is $\mu=55$ cm$^2$/Vs, whereas the scattering on ionized impurities limits the mobility to $\mu=62$ cm$^2$/Vs for temperatures lower than 180 K. The Seebeck coefficient relative to aluminum at room temperature is $S_{\text{ZnGa}_2\text{O}_4-\text{Al}}=(-125\pm2)\;\mu$V/K and shows a temperature dependence as expected for degenerate semiconductors. At low temperatures, around 60 K we observed a maximum of the Seebeck coefficient due to the phonon drag effect. 
\end{abstract}

\maketitle

Transparent conducting oxides (TCOs) have drawn attention due to their possible application in high power, optical or gas sensing devices\cite{Grundmann2010,Lorenz2016,Fortunato2012,Higashiwaki2016,Suzuki2007,Green2016,Chabak2016,Galazka2018}. Recently, $\beta$-Ga$_2$O$_3$ and related semiconducting oxides with ultra-wide bandgaps of over 4 eV are in the focus, since they offer transparency in the visible spectrum, semiconducting behaviour and breakthrough electric fields of several MV/cm. The fundamental research has been extended from binary to ternary and quaternary systems to find new substrate material for epitaxial thin film growth, as well as to make use of a higher degree of freedom in terms of doping\cite{Galazka2019}.\\
ZnGa$_2$O$_4$ is a novel ternary conducting oxide that crystallizes in the spinel crystal structure, which makes it interesting as a substrate for ferrite spinels\cite{Galazka2019}. Furthermore, the material might be promising for electric application, which gives rise to a study of the fundamental electric and thermoelectric transport properties. The isotropic thermal conductivity at room temperature is $\lambda=22$ W/mK\cite{Galazka2019}, but many other material parameters remain to be clarified. Theoretical values for the bandgap of ZnGa$_2$O$_4$ were predicted to be indirect (K-$\Gamma$) with values between \hbox{2.69 eV} - \hbox{4.71 eV}\cite{Zerarga2011,Pisani2006,Lopez2009,Karazhanov2010,Brik2010,Dixit2011,Xia2018}, or direct ($\Gamma$-$\Gamma$) with values of 2.79 eV\cite{Sampath1999}. The experimental bandgap was found at the values of \hbox{4.0 eV} – \hbox{5.0 eV} measured on synthetized ZnO:Ga$_2$O$_3$ powders\cite{Sampath2005}, ceramics\cite{Omata1994}, films obtained by mist-CVD\cite{Oshima2014}, films obtained by sol-gel\cite{Zhang2010} and on bulk crystals obtained by the flux method\cite{Yan2005}, while experimental optical bandgap measured on bulk single crystals obtained from the melt is 4.6 eV\cite{Galazka2019}. Theoretical calculated effective masses are in the range of $m^*=0.22-0.66\;m_\text{e}$\cite{Sampath1999,Karazhanov2010,Dixit2011,Xia2018}. Only little is known about electric transport parameters. On one hand, ZnGa$_2$O$_4$ ceramics show low electrical conductivity \hbox{($\sigma_{\text{ZnGa}_2\text{O}_{4,\text{ceram.}}}\approx 30$ S/cm \cite{Omata1994})}. On the other hand, melt-grown bulk single crystals showed high electrical conductivity of about 50 - 500 S/cm\cite{Galazka2019}.\\
In this work, we investigate as-grown bulk ZnGa$_2$O$_4$ of blueish coloration
and perform temperature-dependent Seebeck-, van-der-Pauw- and Hall-measurements between $T=10$ K and $T=310$ K. We discuss the results in terms of electron scattering processes and thermoelectric effects observed in a degenerate semiconductor.\\  
The samples have been grown using the vertical gradient freeze method\cite{Galazka2019}, without intentional doping. Powdered ZnO and Ga$_2$O$_3$, each with a purity of 99.999 \% were used as starting materials. The growth was performed in an Ir crucible with an excess of 0.2 mol.\% of ZnO to compensate its losses at high temperatures. The growth atmosphere consisted of O$_2$/Ar (20 vol.\% / 80 vol.\%). The heating up and cooling down times were 7 and 10 hours, respectively, and resulted in single crystal blocks with blue and transparent appearance. As can be seen from ref. \cite{Galazka2019} \hbox{Fig. 7}, the composition of the obtained crystals were close to stoichiometric. All crystals grown in this way showed similar defect structure as has been reported\cite{Galazka2019}, without low-angle grain boundaries. For further details on the growth conditions, structural and chemical characterization of the material, see Galazka, \textit{et al.}\cite{Galazka2019}.\\
\begin{figure}[!h]
\includegraphics[scale=1]{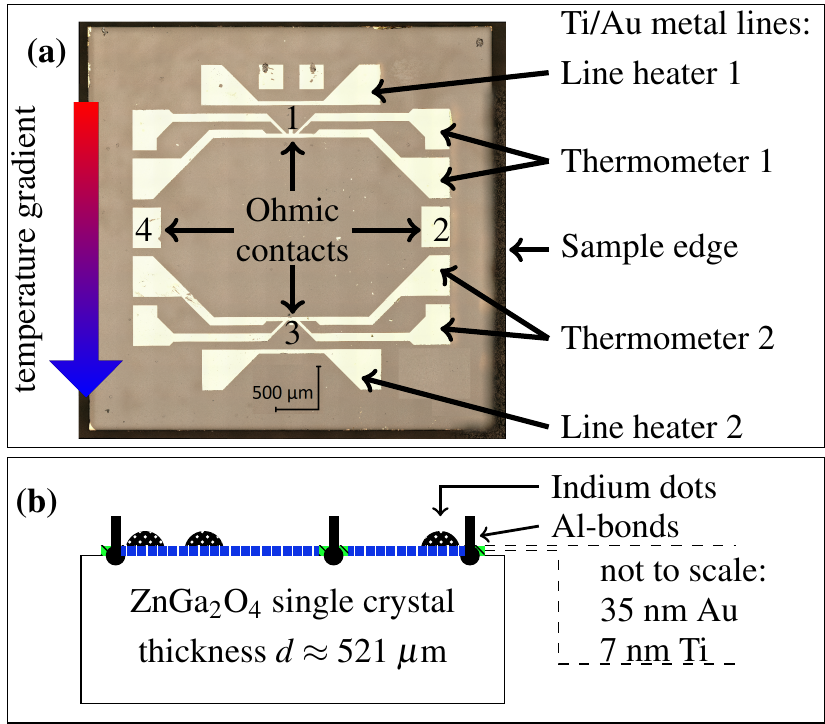} 
\caption{Figure (a) gives an overview of the function of the Seebeck micro lab consisting of Ti/Au lines. A two point conductor at the top (or bottom) of the picture serves as a line heater (\textit{Line heater 1}) to create a temperature gradient $\Delta T$ across the sample. The temperature difference is being measured by the change of the four point resistance in the thermometer lines at the bottom of the sketch (\textit{Thermometer 2}) and below the line heater (\textit{Thermometer 1}). At the marks 1 and 3 Ohmic contacts were prepared, which are used to measure the thermo voltage ($U_\text{th}$). Additional Ohmic contacts (2 and 4) are located at the sample edges and allow van-der-Pauw measurements.  Figure (b) illustrates the cross-section of the sample, see text for details.}\label{Scheme}
\end{figure}
To perform electric and thermoelectric measurements, a microlab, consisting of metal lines, has been manufactured on the surface. \hbox{Figure \ref{Scheme}} shows a microscopic image and a schematic view of the microlab, which allows the measurement of the Seebeck coefficient, conductivity using the van-der-Pauw method, as well as the Hall resistance. The Ohmic contacts used for measuring the thermovoltage $U_\text{th}$ are located in the middle of the four-point metal lines (Ohmic contacts 1 and 3 in \hbox{Fig. \ref{Scheme}}), which serve as thermometers and allow the measurement of the temperature difference $\Delta T$.
\\
The microlab has been manufactured by standard photolithography and magnetron sputtering of titanium (7 nm) and gold (35 nm), after cleaning with acetone and isopropanol and subsequent drying. The as-sputtered metal lines of the microlab are isolated due to a Schottky contact relative to the ZnGa$_2$O$_4$ bulk crystal. Ohmic contacts with the ZnGa$_2$O$_4$ bulk crystal were achieved by direct wedge bonding with an Al/Si-wire (99\%/1\%) on the deposited metal structure, creating point contacts. To keep some parts of the microlab isolated relative to the thin film, the electrical contacts were prepared by attaching gold wire with indium to the Ti/Au metal lines. This procedure can be compared to the one used with $\beta$-Ga$_2$O$_3$, see\cite{Ahrling2019,Boy2019}. \hbox{Figure \ref{I-V_curves}} displays exemplary two-point I-V curves at room temperature.\\
The experimental procedure is carried out in a flow cryostat between $T=10$ K and $T=320$ K. After the bath temperature is stabilized, the van-der-Pauw and Hall measurements are carried out. Subsequently the Seebeck measurements are performed. The Seebeck measurements involve the creation of various temperature differences by imprinting different currents into the line heater. The thermovoltage is measured simultaneously for approximately three minutes, which allows to create a stable temperature difference across the sample. Then, while keeping the heating current constant, the resistances of the thermometers are being measured. This procedure is repeated within bath temperature intervals of 10 K.\\
In the following, we present the measurement results of the electric and the thermoelectric transport, as shown in \hbox{figs. 2-8}.
\begin{figure}[!]
\includegraphics[scale=1]{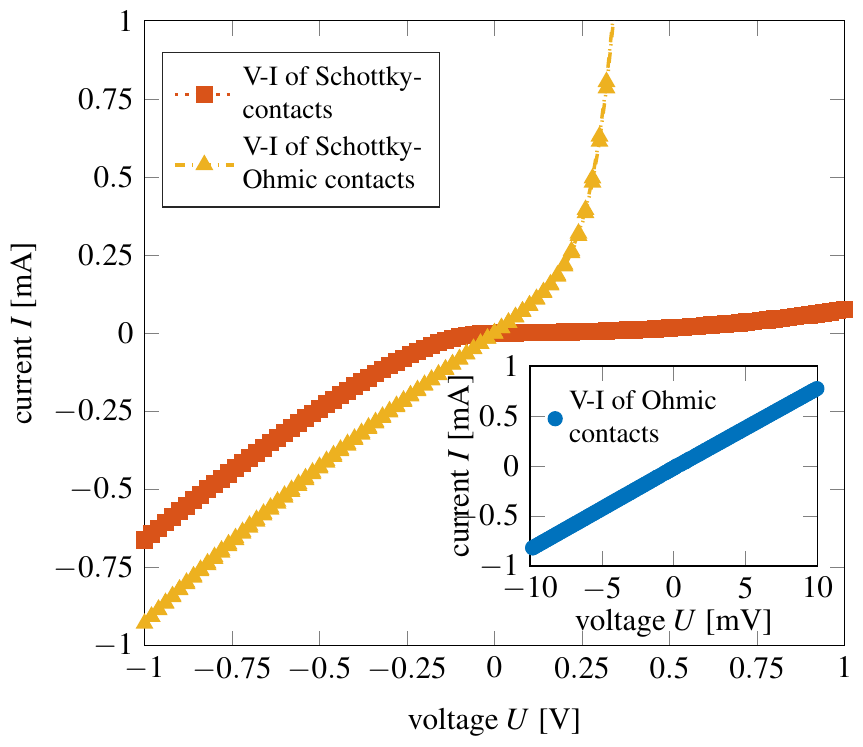} 
\caption{Exemplary two-point $I-V$ curves at room temperature between two Ohmic contacts archieved by wedge-bonding with aluminum wire (inset) with a resistance of $R_\text{ohm.}\approx 12\;\Omega$, two seperated heater lines manufactured by standard photolithography and magnetron sputtering of Ti/Au (7 nm/35 nm) which have been contacted with Au wire and Indium (red squares) and a Schottky diode contact (yellow triangles). The dynamic resistance of the Schottky barriers in reverse bias is at least two orders of magnitude higher $R_\text{dyn.,sch}\geq 1000\;\Omega$.
} \label{I-V_curves}
\end{figure}
To quantify the quality of the contacts, two-point $I-V$ curves were measured at room temperature and are shown in \hbox{Fig. \ref{I-V_curves}}. The red square and orange triangles in the main plot show the $I-V$ curves of Schottky-Schottky and Schottky-Ohmic contacts, respectively.  The $I-V$ curve for the Ohmic-Ohmic contact configuration can be seen as blue circles in the inset. The two-point $I-V$ graphs of the different Schottky contact configurations show the expected diode curve for forward bias (positive voltages) as well as the reverse bias (negative  voltages). The dynamic resistance of the Schottky-Ohmic and Schottky-Schottky contacts are at least two orders of magnitude higher than the Ohmic-Ohmic contact resistance of \hbox{12 $\Omega$} shown in the inset. Therefore a good electrical isolation between the heater lines and the ZnGa$_2$O$_4$ bulk crystal is concluded.\\
\begin{figure}[h!]
\includegraphics[scale=1]{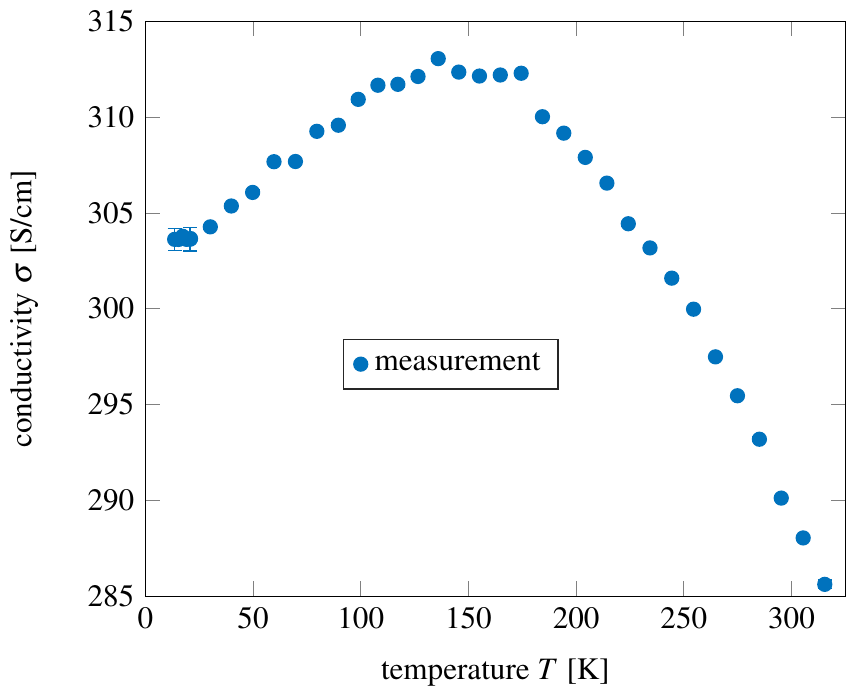} 
\caption{Conductivity $\sigma$ as a function of temperature $T$. The conductivity reaches a maximum in the temperature range of $T=125$ K to $T=175$ K and shows a rather weak temperature dependence.}\label{sigma_vs_T}
\end{figure}
The temperature dependence of the electrical conductivity $\sigma$ is shown in \hbox{Fig. \ref{sigma_vs_T}}. The conductivity is between $285$ and $315$ S/cm for the entire temperature range. For higher temperatures $T\geq260$ K the conductivity $\sigma$ is decreasing. A maximum can be identified around $T=150$ K.\\
\begin{figure}[!]
\includegraphics[scale=1]{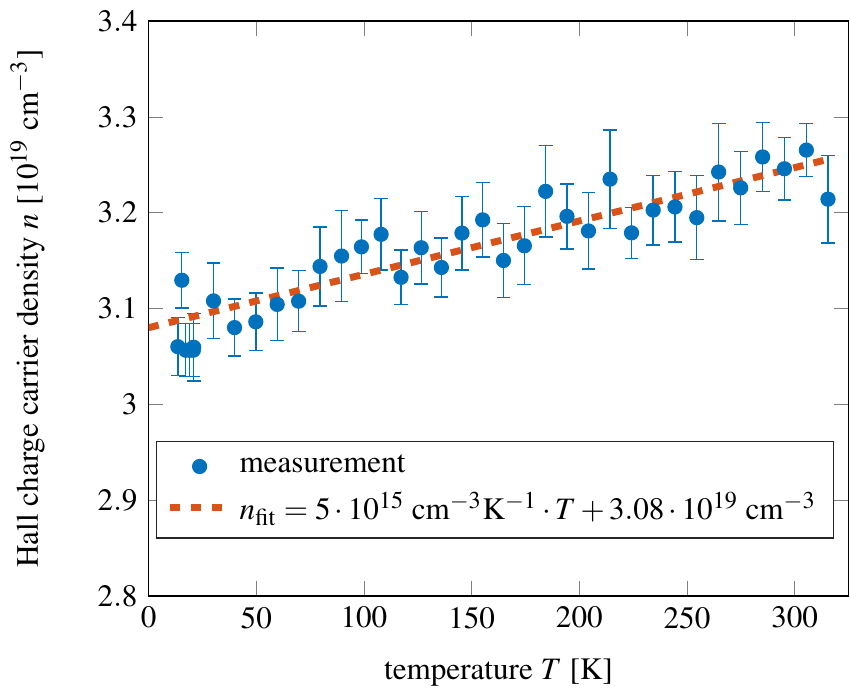} 
\caption{Hall charge carrier density $n$ as a function of temperature $T$. A simple linear fit model has been applied to the data (red dashed line). It can be seen, that the overall change of charge carrier density in the temperature range of $T=10$ K to $T=320$ K is more than one order of magnitude smaller than the charge carrier density.} \label{n_vs_T}
\end{figure}
Hall measurements were performed to determine the charge carrier density, which is depicted as a function of temperature in \hbox{Fig. \ref{n_vs_T}}. A linear fit has been added to the plot, showing the weak temperature dependence. The Hall charge carrier density is in the range of $n=3.1\cdot10^{19}$ cm$^{-3}$ for the entire temperature range.\\
\begin{figure}[!]
\includegraphics[scale=1]{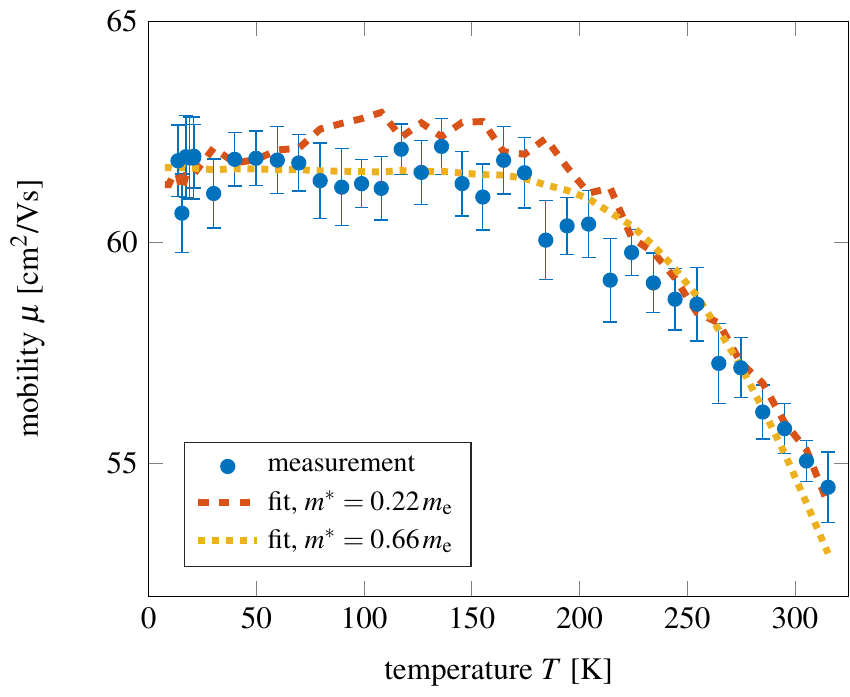} 
\caption{Mobility $\mu$ as a function of temperature $T$. Two fits for $m^*=0.22\,m_\text{e}$ (red dashed line) and $m^*=0.66\,m_\text{e}$ (yellow dotted line) have been calulated and plotted. The high temperature dependence can be explained by electrons scattering with polar optical phonons. One can observe a saturation of the mobility for temperatures lower than $T\leq180$ K. This can be explained by electrons scattering with ionized impurities.
}\label{mu_vs_T}
\end{figure}
The measurement of the electrical conductivity and Hall charge carrier density allows the calculation of the electron mobility $\mu$. The electron mobility as a function of temperature is plotted in \hbox{Fig. \ref{mu_vs_T}}. At room temperature a mobility of $\mu=55$ cm$^2$/Vs was measured. The mobility increases with decreasing temperature, until it reaches a plateau-like feature of $\mu=62$ cm$^2$/Vs at $T=180$ K and below.\\
\begin{figure}[!]
\includegraphics[scale=1]{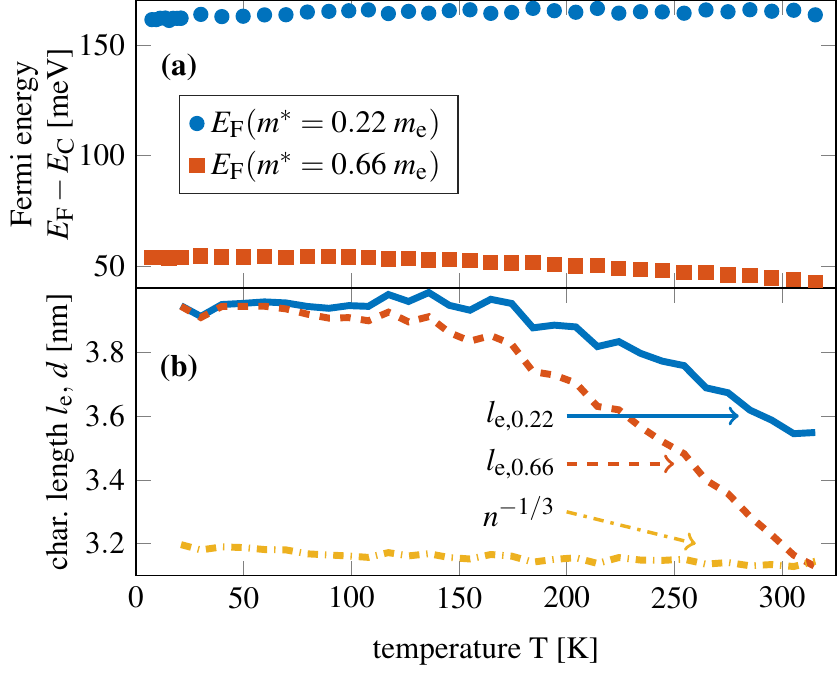} 
\caption{a) The calculated (Eq. \eqref{equation_eta}) relative Fermi energy for effective masses of $m^*=0.22\;m_\text{e}$ (blue dots) and $m^*=0.66\;m_\text{e}$ (red squares) as a function of temperature is shown. The Fermi level lies in the conduction band at all temperatures and shows only weak temperature dependence. b) The calculated mean free path $l_{\text{e,} m^*/m_\text{e}}$ of the electrons as a function of temperature $T$ for $m^*=0.22m_\text{e}$ (blue solid line) and $m^*=0.66m_\text{e}$ (red dashed line), as well as the mean free distance between single ionized donors (yellow dashdotted line), assuming a simple cubic distribution, is shown.}\label{l_e_vs_T}
\end{figure}
To understand the results of the mobility, we calculated the mean free path $l_\text{e}$ of the electrons as a function of temperature, shown in \hbox{Fig. \ref{l_e_vs_T} b)}. The mean free path can be calculated with the following formula
\begin{equation}
l_\text{e}=\sqrt{2E_\text{F}m^*}\frac{\mu}{e}.
\end{equation}
Here, $E_\text{F}$ is the Fermi energy and $e$ is the elemental charge. The Fermi energy was computed from the reduced electron chemical potential $\eta=E_\text{F}/k_\text{B}T$ with the Boltzmann constant $k_\text{B}$. The reduced electron chemical potential $\eta$ was calculated after Nilsson\cite{Nilsson1973}. This method interpolates the range between non-degenerated and degenerated semiconductors and determines the reduced electron chemical potential $\eta$ as follows
\begin{equation}\label{equation_eta}
\eta=\frac{\ln\frac{n}{N_\text{C}}}{1-\left(\frac{n}{N_\text{C}}\right)^2}+\nu\left(1-\frac{1}{1+(0.24+1.08\nu)^2}\right),
\end{equation}
\begin{equation}
\nu=\left(\frac{3\sqrt{\pi}\frac{n}{N_\text{C}}}{4}\right)^{2/3},
\end{equation}
with $N_\text{C}$ being the effective density of states in the conduction band,\\
\begin{equation}
N_\text{C}=2\left(\frac{2\pi m^* k_\text{B}T}{h^2}\right)^{3/2},
\end{equation}
with $h$ the Planck's constant. For effective masses of $m^*=0.22\;m_\text{e}$ and $m^*=0.66\;m_\text{e}$ Fermi energies of $160$ meV and $50$ meV above the conduction band minimum were obtained, respectively. This can be seen in \hbox{Fig. \ref{l_e_vs_T} a)}.\\
\begin{figure}[!]
\includegraphics[scale=1]{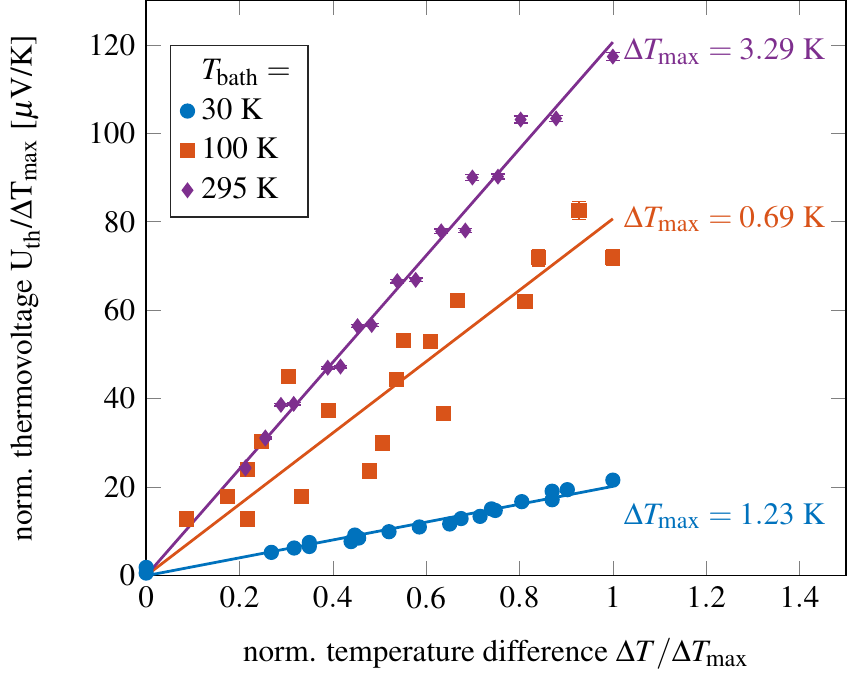} 
\caption{The measured thermovoltage $U_\text{th}$ normalized by the maximum of the temperature difference at different bath temperatures as a function of the normalized temperature difference. An offset $U_\text{os}<50\;\mu$V has been subtracted from the plotted data.}\label{Uth_vs_dT}
\end{figure}
In order to determine the thermoelectric properties, the thermovoltage is measured as a function of temperature difference. \hbox{Figure \ref{Uth_vs_dT}} shows the normalized thermovoltage as a function of normalized temperature difference for bath temperatures of 30 K, 100 K, and 295 K. Small offsets ($U_\text{os}<50\;\mu$V) have been substracted from the data. The Seebeck coefficient is determined by
\begin{equation}
S=-\frac{U_\text{th}}{\Delta T}.
\end{equation}
The change of the Seebeck coefficient as a function of bath temperature can be observed by the change of slope of the linear fits. Furthermore the maximum achieved temperature difference $\Delta T_\text{max}$ is depicted in \hbox{Fig. \ref{Uth_vs_dT}}.\\
\begin{figure}[!]
\includegraphics[scale=1]{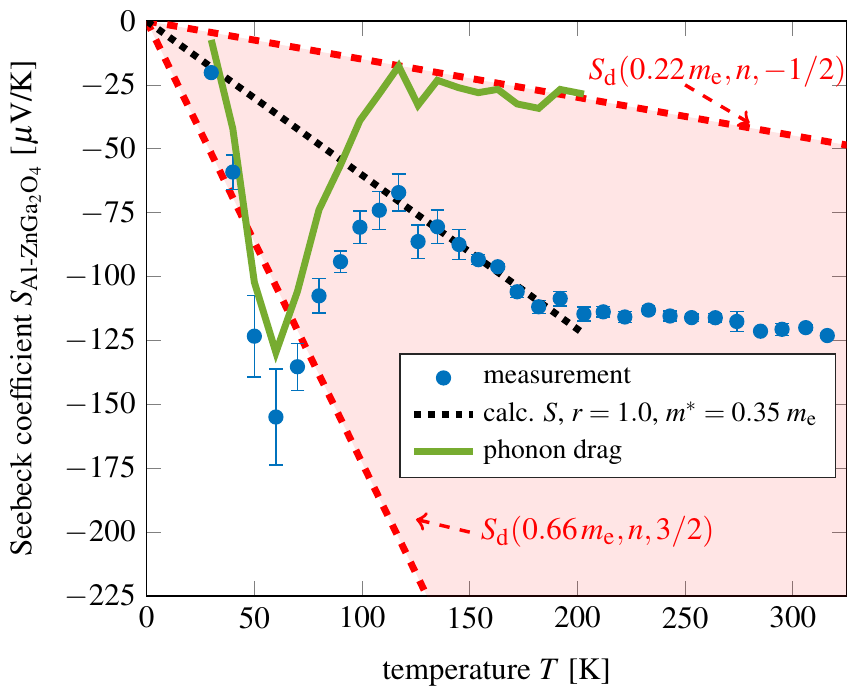} 
\caption{Seebeck coefficient $S$ as a function of temperature $T$. A red area marks the values for the Seebeck coefficients, that can be calculated with equation \eqref{S_d_formula} if the effective mass and scattering parameter are varried between $0.22-0.66\,m_\text{e}$ and $-1/2-3/2$, respectively.
For $T\leq200$ K a calculated Seebeck coefficient with $r=1.0$ and $m^*=0.35\,m_\text{e}$ has been plotted (black dotted line), which fits the data well for \hbox{$100$ K$\leq T\leq200$ K}.
Furthermore, for $T\leq200$ K the difference between the measured and calculated Seebeck coefficient has been plotted and is assumed to be due to the phonon drag effect (green solid line).} \label{S_vs_T}
\end{figure}
The Seebeck coefficient $S$ has been determined for temperatures between 30 K and 320 K. The results are shown in \hbox{Fig. \ref{S_vs_T}} as a function of temperature $T$. The Seebeck coefficients are in the range of $S=-25\;\mu$V/K at $T=30$ K to $S=-125\;\mu$V/K at room temperature. Below room temperature, the magnitude of the Seebeck coefficient decreases steadily down to \hbox{$T=100$ K}. For lower temperatures it shows a maximum at $T=60$ K.\\\\
In the following, we discuss all electrical and thermoelectrical properties in detail.
The electrical conductivity shown in \hbox{Fig. \ref{sigma_vs_T}} has a weak temperature dependence when compared with $\beta$-Ga$_2$O$_3$\cite{Ahrling2019}. As can be seen, this originates partly from the very weak temperature dependence of the Hall charge carrier density shown in \hbox{Fig. \ref{n_vs_T}}. The unintentionally doped ZnGa$_2$O$_4$ is a degenerate semiconductor, which we conclude from the high magnitude and the weak temperature dependence of the charge carrier density and the calculated relative Fermi-level shown in \hbox{Fig. \ref{l_e_vs_T} a)}, which is above the lower conduction band edge for all cases. Furthermore, this is in agreement with the Mott-criterium, which gives an approximation for a critical charge carrier density $n_\text{c}=\left(\frac{m^*e^2}{16\pi\epsilon_0\epsilon_\text{s}\hbar^2}\right)^3$ above which the semiconductor is degenerate. $\epsilon_0$ is the vacuum permittivity and $\hbar$ is the reduced Planck's constant. For the static dielectric constant $\epsilon_\text{s}=9.88$ has been used, which was obtained for Zn$_{0.99}$Cu$_{0.01}$Ga$_2$O$_4$\cite{Lu2017}. The critical charge carrier density is $n_\text{c}(m^*=0.22\cdot m_\text{e})=1.2\cdot10^{18}$ cm$^{-3}$ and $n_\text{c}(m^*=0.66\cdot m_\text{e})=3.1\cdot10^{19}$ cm$^{-3}$, depending on the assumed effective masses $m^*=0.22-0.66\;m_\text{e}$. The critical carrier density lies underneath the Hall charge carrier density for all cases. \\
The charge carrier mobility was calculated and shown in \hbox{Fig. \ref{mu_vs_T}}. For the high temperature regime, it is limited by optical phonon scattering (OP). OP scattering can deviate significantly from the $T^{-3/2}$ dependence of acoustic deformation potential scattering\cite{Seeger2004} due to its inelastic nature.\\
In the chemically related TCOs $\beta-\text{Ga}_2\text{O}_3$\cite{Ma2016} and ZnO\cite{Bikowski2014} it was shown that polar optical phonon scattering is the dominant scattering mechanism at high temperatures due to the partial ionic bonding. The following model after Askerov has proved to be useful for the interpretation of polar optical phonon scattering in degenerate semiconductors\cite{Askerov1994,Bikowski2014}\\
\begin{equation}\label{eq:muPOP}
\mu_\text{POP,deg}=\frac{e\hbar}{4\alpha E_\text{POP}m^*}\left(e^{\frac{E_\text{POP}}{k_\text{B}T}}+1\right)
\end{equation}
with\\
\begin{equation}
\alpha=\left(\frac{1}{\epsilon_\infty}-\frac{1}{\epsilon_s}\right)\frac{\sqrt{m^*}e^2}{4\pi\epsilon_0\hbar\sqrt{2E_\text{POP}}}.
\end{equation}
$E_\text{POP}$  the average energy of the optical phonons, $\alpha$ the polaron coupling constant and $\epsilon_\infty$ is the high frequency dielectric constant. The high frequency dielectric constant of $\beta-\text{Ga}_2\text{O}_3$\cite{Passlack1994,Rebien2002} ($\epsilon_\infty=3.57$) has been used since the exact value for $\text{ZnGa}_2\text{O}_4$ is unknown and it is expected in the same range as the values for ZnO\cite{Ashkenov2003,Decremps2002} ($\epsilon_\infty\approx3.7$).\\
The mean free path in \hbox{Fig. \ref{l_e_vs_T} b)} shows, that there is a temperature independent process that limits the mean free path at $l_\text{e,max}=4$ nm. This limit becomes clear for temperatures lower than $T\leq 180$ K when electron-phonon interaction becomes weaker. Furthermore, the mean free distance between single ionized donors, assuming a simple cubic distribution, is shown. One can see, that there is an upper limit of $l_\text{e,max}=4$ nm and that the simple model for the donor distribution predicts a mean distance in the same range.\\
There are two approaches to explain the low temperature limit of the electron mobility. On one hand, it can be explained by the scattering of electrons with neutral impurities. The Hall charge carrier data in \hbox{Fig. \ref{n_vs_T}} suggests a constant ionization of the donors, acceptors and vacancies for the investigated temperature interval, so $N_\text{II}=const.$ This leads to the assumption, that also the neutral impurity density $N_\text{NI}=const.$ Electron scattering on neutral impurities can be described by\cite{Seeger2004} $\mu_\text{NI}\;\propto\;N_\text{NI}^{-1}$. Thus, if there is no change in concentration of the neutral impurities, there will be a temperature independent upper limit of the mobility, which can be observed in \hbox{Fig. \ref{mu_vs_T}}.\\
On the other hand, high electron concentrations in semiconductors mean, that there is either a high concentration of singly ionized impurities, or a lower density of ionized impurities with a higher degree of ionization. Furthermore, the scattering of electrons in degenerate semiconductors with ionized impurities can be described by the Brooks-Hering equation\cite{Bikowski2014,Dingle1955,Ellmer2001}\\
\begin{equation}\label{eq:muII}
\mu_\text{II,deg}=\frac{n}{Z^2N_\text{II}}\frac{24\pi^3(\epsilon_0\epsilon_\text{s})^2\hbar^3}{e^3m^{*^2}}\frac{1}{\ln[1+\beta(n)]-\frac{\beta(n)}{1+\beta(n)}}
\end{equation}
with\\
\begin{equation}
\beta(n)=\frac{3^{1/3}4\pi^{8/3}\epsilon_0\epsilon_\text{s}\hbar^2n^{1/3}}{e^2m^*}.
\end{equation}
$Z$ is the degree of ionization and $N_\text{II}=N_\text{D}+N_\text{A}=2N_\text{A}+n$ is the density of ionized impurities, with the donator and acceptor densities $N_\text{D}$ and $N_\text{A}$, respectively. The ionized impurity scattering is expected to be more dominant, since there is a high density of ionized donors due to the high Hall charge carrier density.\\
The fits shown in \hbox{Fig. \ref{mu_vs_T}} have been calculated using the Matthiessen's rule and consider the scattering on polar optical phonons and ionized impurities after eq. \eqref{eq:muPOP} and eq. \eqref{eq:muII}, respectively 
\begin{equation}
\mu=\left(\frac{1}{\mu_\text{POP,deg}}+\frac{1}{\mu_\text{II,deg}}\right)^{-1}
\end{equation}
for effective masses of $m^*=0.22\,m_\text{e}$ and $m^*=0.66\,m_\text{e}$. The $m^*=0.22\,m_\text{e}$ fit (red dashed line) results in an acceptor density $N_\text{A}=6.65\cdot10^{19}$ cm$^{-3}$ and a phonon energy of $E_\text{POP}=87$ meV. For the $m^*=0.66\,m_\text{e}$ fit (yellow dotted line) an acceptor density $N_\text{A}=6.15\cdot10^{18}$ cm$^{-3}$ and a phonon energy of $E_\text{POP}=140$ meV were obtained. These results lead to compensation ratios ($N_\text{A}/N_\text{D}$) between $13\,\%$ and $70\,\%$ for $m^*=0.66\,m_\text{e}$ and $m^*=0.22\,m_\text{e}$, respectively.\\
The measurement of the thermovoltage as a function of temperature difference, as shown in \hbox{Fig. \ref{Uth_vs_dT}}, reveals that there is a major change in the maximum reached temperature difference for different bath temperatures. This also correlates with the precision of the data points. The maximum reached temperature difference depends on the imprinted heating power, but is more strongly dependent on the thermal conductivity of the material. The higher the thermal conductivity of the material, the more difficult it is to create large temperature differences. This is the main reason for the increasing uncertainty of the Seebeck coefficient going to lower bath temperatures. From the change of maximum temperature difference, one can conclude the change of the thermal conductivity of the material. Having a look at the precision of the data in \hbox{Fig. \ref{S_vs_T}}, which is correlated with the maximum temperature difference and therefore with the thermal conductivity, one can see, that the thermal conductivity seems to have a maximum around $T_\text{bath}=60$ K and decreases as the temperature decreases further. This could be due to a distortion of the lattice, which has been reported in \cite{Galazka2019} as particles revealing Moir\'e patterns in transmission electron microscopy bright field images.\\
The Seebeck coefficient (\hbox{Fig. \ref{S_vs_T}}) is negative, which means that electrons are the majority charge carriers. This is in agreement with the Hall charge carrier results. The Seebeck coefficient is lower than the one reported\cite{Boy2019} for $\beta$-Ga$_2$O$_3$ in the same temperature regime. This can be understood, since the semiconducting oxide investigated here is degenerate.\\
The red area in \hbox{Fig. \ref{S_vs_T}} marks calculated Seebeck coefficients following the commonly used equation\cite{Seeger2004} for degenerate semiconductors assuming the effective mass to be between $m^*=0.22\;m_\text{e}$ and $m^*=0.66\;m_\text{e}$ and the scattering factor $r$ to be between $r=-0.5$ and $r=1.5$\\
\begin{equation}\label{S_d_formula}
S_\text{d}(m^*,n,r)=-\frac{k_\text{B}}{e}\left(r+\frac{3}{2}\right)\frac{\pi^2}{3}\frac{1}{\eta}.
\end{equation}
The scattering parameter $r$ is based on the assumption, that the electron relaxation time $\tau_\text{e}$ follows a power law dependence $\tau_\text{e}\propto E^r$. In other investigations it was observed, that $\mu\propto E^{r'}$ and $r\approxeq r'+1$ holds in the investigated temperature interval. In general $\mu=\mu(E,T)$ and $r'$ is calculated by the assumption $E=E(T)=k_\text{B}T$ and
\begin{equation}
\frac{\text{d}\ln(\mu)}{\text{d}\ln(T)}=r'+\frac{\partial r}{\partial\ln(T)}.
\end{equation}
A calculated Seebeck coefficient with $r=1.0$ and $m^*=0.35\,m_\text{e}$ was added as a thin black dashed line, which fits the data for \hbox{$100$ K$\leq T\leq200$ K} best. For $T<200$ K the difference of the calculated Seebeck coefficient with $r=1.0$, $m^*=0.35\,m_\text{e}$ and the measured Seebeck coefficient is plotted as a green solid line. We account the observed deviation of the theoretical Seebeck coefficient to be due to the phonon drag effect.\\
In conclusion, we have shown, that as-grown, bulk \hbox{ZnGa$_2$O$_4$} single crystals show higher electrical conductivity at room temperature \hbox{($\sigma_{\text{ZnGa}_2\text{O}_4}\approx 300$ S/cm}) than in earlier investigated ZnGa$_2$O$_4$ ceramics \hbox{($\sigma_{\text{ZnGa}_2\text{O}_{4,\text{ceram.}}}\approx 30$ S/cm \cite{Omata1994})}, as-grown $\beta-$Ga$_2$O$_3$ bulk \hbox{($\sigma_{\beta-\text{Ga}_2\text{O}_3}\approx 3$ S/cm\cite{Ahrling2019})} or as-grown ZnO bulk \hbox{($\sigma_{\text{ZnO}}\approx 40$ S/cm\cite{Xin-Hua2006})} due to the high charge carrier density. The donor mechanisms remain to be established. The wide band-gap of the material makes it suitable for application in high-power devices, which can become even more promising, if $p$-doped material becomes available.
In terms of the power factor $Pf=\sigma S^2$ for thermoelectric applications, ZnGa$_2$O$_4$ has a room temperature value of \hbox{$Pf_{\text{ZnGa}_2\text{O}_4}\approx4.7\;\mu$W/K,} being more than 5 times higher than that of $\beta-$Ga$_2$O$_3$ with \hbox{$Pf_{\beta-\text{Ga}_2\text{O}_3}\approx0.8\;\mu$W/K}. Therefore, ZnGa$_2$O$_4$ is a more promising material for thermoelectric applications of transparent conducting oxides.

\section*{Acknowledgement}
This work was performed in the framework of GraFOx, a Leibniz-ScienceCampus partially funded by the Leibniz association and by the German Science Foundation (DFG-FI932/10-1 and DFG-FI932/11-1).

\nocite{*}
\bibliography{mybibfile}

\end{document}